\documentclass[a4paper]{cas-dc}

\usepackage[numbers,sort&compress]{natbib}
\usepackage[caption=false,font=footnotesize]{subfig}
\usepackage{caption}
\usepackage{threeparttable}
\def\tsc#1{\csdef{#1}{\textsc{\lowercase{#1}}\xspace}}
\tsc{WGM}
\tsc{QE}
\tsc{EP}
\tsc{PMS}
\tsc{BEC}
\tsc{DE}

\begin{document}
\let\WriteBookmarks\relax
\def\floatpagepagefraction{1}
\def\textpagefraction{.001}
\shorttitle{}   
\shortauthors{Ying Zang et~al.}

\title [mode = title]{Deep3DSketch+: Obtaining Customized 3D Model by Single Free-Hand Sketch through Deep Learning}                      




\author[1]{Ying Zang}

\fnmark[1]


\address[1]{College of Information Engineering, Huzhou University, Huzhou 313000, China.}
\address[2]{College of Computer Science and Technology, Zhejiang University, Hangzhou 310027, China.}
\address[3]{KOKONI, Moxin (Huzhou) Technology Co., LTD. Huzhou 313000, China.}

\author[1]{Chenglong Fu}
\fnmark[1]



\author[2,3]{Tianrun Chen}
\cormark[1]
\fnmark[1]
\ead{Tianrun.chen@zju.edu.cn}


\cortext[4]{* Corresponding author}

\nonumnote{First Author, Second Author and Third Author contribute equally to this work.}
\nonumnote{This work is an extended version of the conference report of Chen, Tianrun, et al., "Deep3DSketch+: Rapid 3D Modeling from Single Free-Hand Sketches", in 2023 International Conference on Multimedia Modeling (pp. 16-28), 2023.}
\author[1]{Yuanqi Hu}
\fnmark[2]
\author[1]{Qingshan Liu}
\fnmark[3]
\author[1]{Wenjun Hu}
\fnmark[4]

\begin{abstract}
As 3D models become critical in today’s manufacturing and product design, conventional 3D modeling approaches based on Computer-Aided Design (CAD) are labor-intensive, time-consuming, and have high demands on the creators. This work aims to introduce an alternative approach to 3D modeling by utilizing free-hand sketches to obtain desired 3D models. We introduce Deep3DSketch+, which is a deep-learning algorithm that takes the input of a single free-hand sketch and produces a com- plete and high-fidelity model that matches the sketch input. The neural network has view- and structural-awareness enabled by a Shape Discriminator (SD) and a Stroke Enhancement Module (SEM), which overcomes the limitations of sparsity and ambiguity of the sketches. The network design also brings high robustness to partial sketch input in industrial applications.Our approach has undergone extensive experiments, demonstrating its state-of-the-art (SOTA) performance on both synthetic and real-world datasets. These results validate the effectiveness and superiority of our method compared to existing techniques. We have demonstrated the conversion of free-hand sketches into physical 3D objects using additive manufacturing. We believe that our approach has the potential to accelerate product design and democratize customized manufacturing. 
\end{abstract}



\maketitle

\section{Introduction}
\vspace*{1\baselineskip}
Today, the use of 3D model has been a central position in many manufacturing processes and product designs \cite{fu2021improved}. The emerging trend towards making CNC machines and 3D printers more readily accessible has resulted in the ability for individuals to even conduct manufacturing processes at home with consumer-grade 3D printers and CNC machines \cite{akinyele2020manufacturing}. The trend calls for tremendous demands for 3D content. 

Traditionally, the generation of 3D content is achieved by manually utilizing Computer-Aided Design (CAD) techniques, a process that is demands a high level of skill and expertise from designers. It requires a sophisticated knowledge of CAD software commands and stra tegies, such as the ability to break down a shape into sequential commands, which can be challenging to achieve \cite{chester2007teaching} not to mention that it is also a labor-intensive and time-consuming process \cite{reddy2018development}.

The limitations inherent in CAD techniques have highlighted the need for alternative methods that offer faster and more user-friendly 3D modeling options. Among these alternatives, sketch-based 3D modeling has emerged as a prom-ising solution in recent years. Sketches serve as a crucial tool in professional design and our everyday lives, as they provide an intuitive means for expressing ideas. This project aims to leverage the advantages of sketches and develop an efficient and intuitive 3D modeling tool that enables users to create models quickly. The resulting models can be seamlessly integrated into manufacturing processes such as 3D printing.

Previous researchers have made several attempts, but thus far, a satisfactory solution has not been achieved. Most existing works either require precise line drawings from multiple views or apply step-by-step workflow with \textbf{strategic knowledge} required \cite{ cohen1999interface, deng2020interactive}, which are not friendly for novice users and also still time-consuming. Other works use template primitives or retrieval-based approaches \cite{chen2003visual, wang2015sketch}, but lack the customizability. To realize the goal of rapid and intuitive 3D modeling, this work propose to use only a single sketch as the input and generates a complete and high-fidelity 3D model. This is a challenging task due to the sparsity and ambiguity of sketches. Sketches are sparse because they have only a single view, are mostly abstract, lack fine boundary information when drawn by humans, and, most critically, lack texture information for depth estimation. This leads to a large amount of uncertainty when learning 3D shapes. The abstract boundary also makes it difficult to interpret as the same set of strokes can have different interpretations in the 3D world, leading to ambiguity. 

\begin{figure*}
\centering
\includegraphics[width=0.9\linewidth]{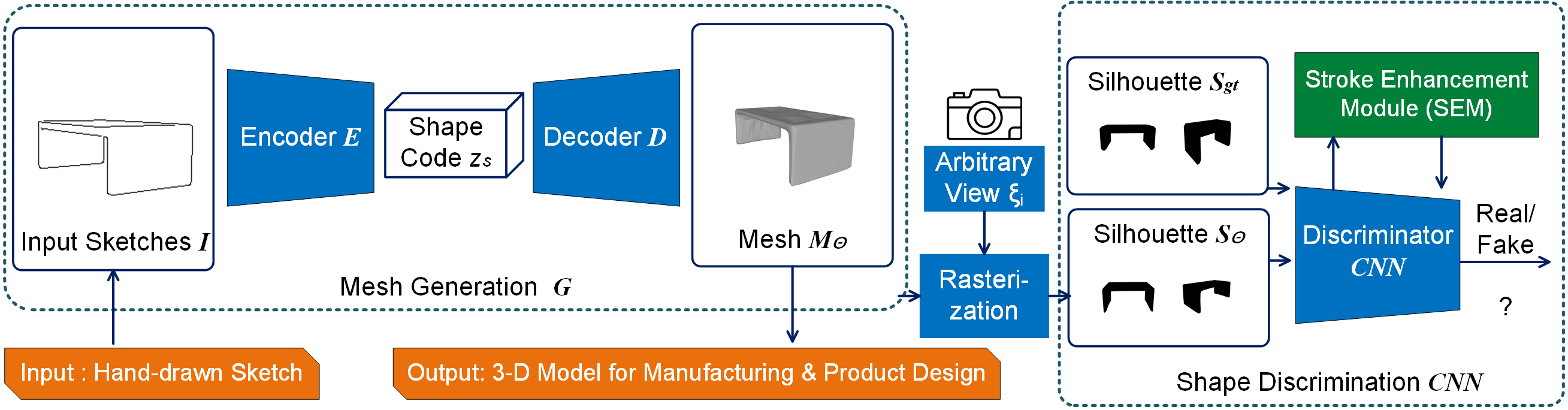}
\caption{\textbf{The overall structure of Deep3DSketch+.} The model consists of a Mesh Generator $G$ and a Shape Discriminator $SD$. The user input a hand-drawn sketch to the Mesh Generator $G$, the neural network produce a Mesh $M_\Theta$ that can be used for manufacturing and product design. The Shape Discriminator (SD) is used to add supervision to the mesh generation.} 
\label{fig:0}
\end{figure*}

To address this challenge, we introduce an innovative end-to-end neural network called \textbf{Deep3DSketch+}. Our proposed solution incorporates a streamlined generation network and a Shape Discriminator (SD) that is sensitive to structural information. By leveraging inputs from both the predicted mesh and ground truth models, our approach significantly enhances the capacity to produce lifelike 3D models.To ensure the watertightness of the generated shape, our approach involves optimizing the offset of a round ball-shaped template based on the user's intention to create the mesh. Furthermore, we introduce a Stroke Enhancement Module (SEM) to enhance the network's ability to extract structural features from sketches and their corresponding silhouettes. By employing the SEM, we improve the network's capability for capturing essential stroke-related information, resulting in more accurate and detailed structural feature extraction. With our novel network, users only need to draw a single-view sketch and the neural network handles the rest. The generated watertight models can be readily exported to manufacturing processes (e.g., the slicer of 3D printers) without repairing non-manifold edges or fixing holes. 

Through extensive experimentation, we have validated the effectiveness of our approach, achieving state-of-the-art (SOTA) performance on both synthetic and real datasets. Our method excels in capturing structure information and generating 3D models with higher fidelity. Furthermore, we conducted experiments to evaluate the robustness of our met-hod to incomplete sketches, a common scenario in industrial applications where drawings may be compromised in dirty environments. The results demonstrated the high resilience of our approach in handling such incomplete sketches.To showcase the practicality and real-world industrial application of our method, we provided an example of exporting the raw output of our sketch-based 3D models for manufacturing using consumer-level 3D printers. This exemplifies the readiness and viability of our approach for practical implementation.Collectively, these results serve as strong evidence of the effectiveness and practical usefulness of our approach, highlighting its potential impact in various industrial applications.

\section{Related Works}
\vspace*{1\baselineskip}
\noindent\textit{2.1. Sketch-based 3D Modeling}
\vspace*{1\baselineskip}

Sketch-based 3D modeling has been a subject of research for many years, with various approaches proposed in the literature \cite{bonnici2019sketch,olsen2009sketch, chen2023deep3dsketch+}.  These approaches can generally be divided into two categories: interactive and end-to-end. Interactive methods involve step-by-step decomposition or specific drawing gestures and annotations \cite{li2020sketch2cad,cohen1999interface,shtof2013geosemantic,deng2020interactive},which require expertise and strategic knowledge. On the other hand, end-to-end approaches, utilizing template primitives or retrieval-based techniques \cite{chen2003visual,wang2015sketch,sangkloy2016sketchy} can generate satisfactory results but lack customization flexibility. Previous research studies \cite{zhang2021sketch2model,guillard2021sketch2mesh}  have utilized deep learning techniques for direct 3D model reconstruction, considering it as a single-view 3D reconstruction task. Nevertheless, it is important to note that sketch-based modeling and traditional monocular 3D reconstruction exhibit notable differences. Sketches are sparse, abstract, and lack textures, requiring additional cues to generate high-quality 3D shapes.In this work, we propose a meth-od that specifically addresses these challenges and provides an efficient and accurate solution for 3D modeling.

\vspace*{1\baselineskip}
\noindent\textit{2.2. Single-View 3D Reconstruction}
\vspace*{1\baselineskip}

The task of single-view 3D reconstruction has been a longstanding challenge in the field of computer vision. The availability of large-scale datasets such as ShapeNet~\cite{chang2015shapenet} has significantly contributed to advancements in data-driven approaches. Some studies~\cite{chen2019learning,park2019deepsdf} leverage category-level information to infer 3D representations from a single image, while others~\cite{liu2019soft,liu2019learning,kato2018neural} directly generate 3D models from 2D images using differentiable rendering techniques. Recent advancements~\cite{lin2020sdf,yu2021pixelnerf} explore unsupervised methods for implicit function representations through differentiable rendering. As for shape representation, most of these works use mesh-based representation for 3D shapes. Unlike other representations \cite{zhang2023dyn,zhang2023painting,fu2022panoptic,dou2020top,dou2022tore,dou2022coverage,xu2022rfeps,lin2023patch,wang2022progressively,yang2023neural}, mesh representation can be directly integrated to existing shape editing tools. However, existing methods predominantly concentrate on learning 3D geometry from 2D images, while our objective is to generate 3D meshes from 2D sketches, which represent a more abstract and sparse representation compared to colored images.Successfully generating high-quality 3D shapes from such abstract sketch representations remains a challenging task that needs to be addressed.

\section{Method}
\vspace*{1\baselineskip}

\noindent\textit{3.1. Preliminary}
\vspace*{1\baselineskip}

The input for 3D modeling is a single binary sketch  - ${I\in \left\{0,1\right\}^{W\times H}}$ , where ${I \left [ i,j \right ] = 0 }$ if marked by the stroke, and ${I\left [ i,j \right ] = 1 }$ otherwise. The network $G$ is specifically designed to generate a mesh ${M_\Theta =(V_\Theta,  F_\Theta)}$, where ${V_\Theta}$ and ${F_\Theta}$ represents the mesh's vertices and facets,respectively. Importantly, the generated mesh ${M_\Theta}$ ensures that its silhouette ${S_\Theta :\mathbb{R}^3} $
$ {\rightarrow \mathbb{R}^2} $ aligns with the input sketch $I$. 

\vspace*{1\baselineskip}
\noindent\textit{3.2. View-Aware and Structure-Aware 3D Modeling}
\vspace*{1\baselineskip}

The method we propose, Deep3DSketch+, is depicted in\ref{fig:0}, showcasing its overall structure.
The network $G$ serving as the backbone, follows an encoder-decoder architecture. As sketches are a sparse and ambiguous form of input, Given that sketches are sparse and inherently ambiguous, the encoder $E$ is responsible for converting the input sketch into a latent shape code $z_s$ This shape code captures the essence of the sketch at a higher level, considering factors such as semantic category and conceptual shape. Then, a decoder $D$ to transfer $z_s$ to the mesh $M_\Theta = D(z_s)$. Instead of using structures like MLP to predict point-wise locations, To obtain the output mesh $M_\Theta$ , we employ cascaded upsampling blocks that progressively infer the 3D shape information at higher spatial resolutions. These upsampling blocks calculate the vertex offsets of a template mesh and deform it accordingly. By gradually increasing the spatial resolution, we refine the shape representation and generate the desired output mesh $M_\Theta$. Such the design ensures that the generated model is both highly detailed and watertight, making it ready for use in manufacturing software such as slicers for 3D printers. 

To supervise the generation process, we render the generated mesh $M_\Theta$ using a differentiable renderer, which produces a silhouette $S_\Theta$. The network is trained in an end-to-end manner, where the supervision is provided by comparing the rendered silhouettes to the ground truth. The gradients of the differentiable renderer are approximated and used to guide the training process, ensuring that the generated mesh aligns with the desired silhouette. 

Despite its effectiveness, the encoder-decoder structured generator \textit{\textbf{G}} encounters challenges in generating high-quality 3D shapes. This can be attributed to the sparse nature of sketches and the limited information available through the single-view silhouette constraint \cite{zhang2021sketch2model,guillard2021sketch2mesh}. Relying solely on these constraints poses limitations in effectively capturing the intricacies of 3D shapes. In order to capture fine details and realistic structures of objects, it becomes necessary to incorporate additional clues into the modeling process. Previous work  \cite{guillard2021sketch2mesh}has proposed a two-stage post-refinement process, wherein a rough shape is initially generated and subsequently optimized to align with the silhouette. This approach allows for the enhancement and refinement of the initial shape, resulting in improved accuracy and realism. However, this meth- od is not efficient enough for real-time applications. In contrast, we aim to develop a more efficient, end-to-end solution that can quickly generate high-fidelity 3D meshes. To accomplish this, we introduce a Shape Discriminator (SD) and a Stroke Enhancement Module (SEM).

\textit{\textbf{1) \ Shape Discriminator (SD) and Multi-view Sampling:}} To tackle this challenge, we propose the introduction of a Shape Discriminator (SD) as part of our approach. During training, the SD incorporates 3D shapes from real datasets to encourage the mesh generator \textit{\textbf{G}} to produce realistic shapes. This addition helps maintain efficiency during inference. To discern the quality and realism of the generated shapes, the discriminator SD takes as input both the generated silhouette derived from the predicted mesh and the rendered silhouette obtained from the manually-designed mesh. By incorporating both silhouettes during the evaluation process, the discriminator SD becomes capable of effectively distinguishing the authenticity and fidelity of the generated shapes. This allows for a more comprehensive assessment of the generated shapes' quality and realism. 

Moreover, we argue that a single silhouette is not enough to encapsulate all the information of the 3D mesh, as it is a 3D shape $M_\Theta$ that can be viewed from different angles, unlike a 2D image translation task. The silhouette constraints ensure that the generated model aligns with the input sketch’s viewpoint,However, incorporating the Shape Discriminator does not inherently ensure that the model produces results that are both realistic and consistent across various viewpoints. To overcome this limitation, we propose to randomly sample $N$ camera poses $\xi_{1...N}$ from a camera pose distribution $p_{\xi}$ to gather more information about the 3D object. The random pose sampling can force the network learns to generate reasonable 3D fine-structured shapes independent from the viewpoints. It is widely acknowledged in the research community that multi-view silhouettes hold valuable geometric information pertaining to the 3D object~\cite{gadelha2019shape,hu2018structure}. The availability of silhouettes from different viewpoints enables a more comprehensive understanding of the object's shape and structure, contributing to more accurate and detailed 3D reconstructions. In our approach, we employ a differentiable rendering module to generate the silhouettes $S_{gt}\left \{1...N \right \} $ from the mesh $M_{gt}$ and the silhouettes $S_{\theta}\left \{1...N \right \} $ from the mesh $M_{\theta}$. The differentiable rendering equation $R$, as described in \cite{liu2019soft}, is utilized to achieve this rendering process. This equation enables us to compute the rendered silhouettes based on the geometry and camera poses, allowing for the integration of the rendering module into the training pipeline. 

By feeding the set of rendered silhouettes  $S_{\theta}\left \{1...N \right \} $  to the Shape Discriminator (SD) for both the predicted meshes and the real meshes, our network becomes aware of the geometric structure of objects across multiple views. This helps ensure that the generated mesh is not only reasonable, but also high-fidelity in terms of details.
\begin{figure}[t]
\includegraphics[width=0.5\textwidth]{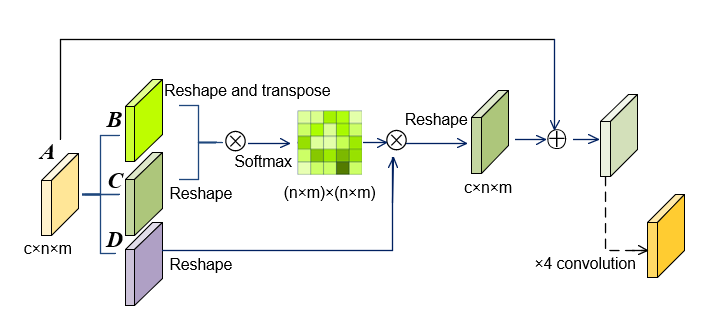}
\caption{{\textbf{The Details of Stroke Enhancement Module (SEM).} {$\otimes$ denotes element-wise multiplication, $\oplus$ demotes element-wise add operation.}}}
\label{fig:3}
\end{figure}

\textit{\textbf{2) \  Stroke Enhancement Module (SEM):}} Sketch-based 3D modeling is distinct from traditional monocular 3D reconstruction due to the limited information provided by sket- ches. While conventional 3D reconstruction relies on rich textures and diverse features present in an image to estimate depth, sketches lack these features and are limited to a single color. This makes it difficult to accurately predict depth in a sketch-based modeling task.Alternatively, we propose the integration of a Stroke Enhancement Module (SEM), as illustrated in Fig. \ref{fig:3}, to fully exploit the monochromatic information for feature extraction. The SEM comprises a position-aware attention module inspired by \cite{fu2019dual}, which incorporates a wide range of contextual information into local features to capture the spatial interdependencies \cite{chen2020supervised}. Additionally, a post-process module is employed to manipulate the features from the position-aware attention module using a series of convolutions. This process smoothly incorporates these enhanced features into the original features before attention in an element-wise manner. This strategy effectively enhances the learning of features in specific positions, particularly along the boundaries.

To elaborate on the implementation, the local feature derived from the silhouette $A \in \mathbb{R}^{C\times N\times M}$ is passed through a convolutional layer to generate two local features, $B, C \in \mathbb{R}^{C\times W}$ where $W=M\times N$ corresponds to the number of pixels. Another convolutional layer is applied to form the feature map $D \in \mathbb{R}^{C\times N\times M}$. After transposing matrix $C$, a matrix multiplication operation is performed with matrix $B$. Subsequently, a softmax layer is applied to the resulting matrix, resulting in the generation of the attention map $S \in \mathbb{R}^{W\times W}$. This attention map effectively enhances the utilization of critical structural information represented by the silhouette.

\begin{equation}
s_{ij} = \frac{exp\left(B_i C_j\right)}{\sum_{i=1}^{W}exp\left(B_i  C_j\right)},
\end{equation}

The attention map is utilized to generate the output $F$ by calculating a weighted sum of the original feature and the features from all positions. This weighted sum is computed based on the attention weights determined by the attention map, 
\begin{equation}
    F_{j} = {\lambda\sum_{i=1}^{W}\left(s_j  D_j\right)+A_j}
\label{pam}
\end{equation}
\begin{figure*}
\centering
\includegraphics[width=0.7\linewidth]{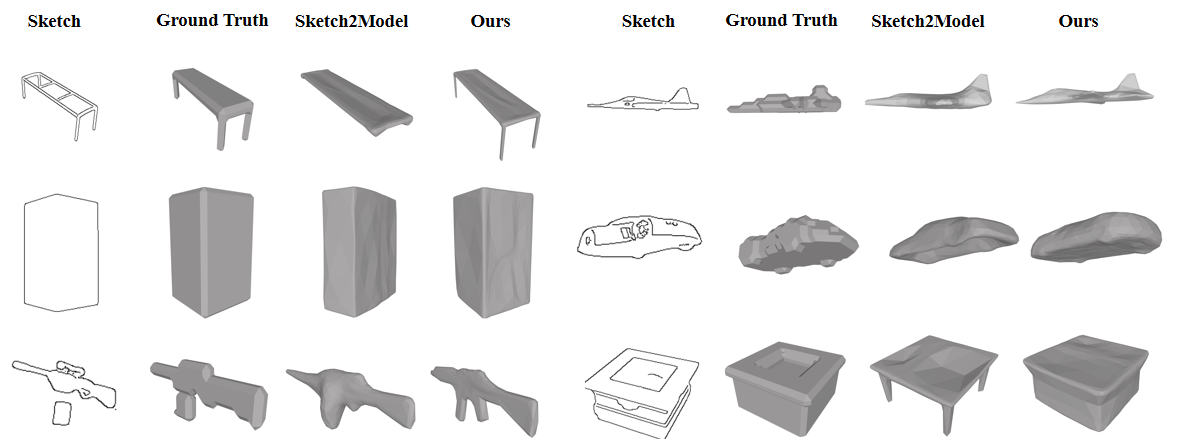}
\caption{{\textbf{Qualitative evaluation with existing state-of-the-art.} The visualization of 3D models generated demonstrated that our approach is capable of obtaining higher fidelity of 3D structures.}}
\label{fig:2}
\end{figure*}
\begin{table*}[width=.72\textwidth, cols=4, pos=h]
\caption{The quantitative evaluation of ShapeNet-Synthetic dataset.}
\begin{center}{
\begin{tabular}{c| ccccccc}
\hline\hline
\multicolumn{8}{c}{Shapenet-synthetic (Voxel IoU $\uparrow)$} \\
\hline
 & car & sofa & airplane & bench & display & chair & table \\
\hline
Retrieval & 0.667 & 0.483 & 0.513 & 0.380  & 0.385 & 0.346 & 0.311 \\

Auto-encoder & 0.769 & 0.613 & 0.576 & 0.467 & 0.541 & 0.496 & \textbf{0.512} \\

Sketch2Model  & 0.751 & 0.622 & 0.624 & 0.481 & \textbf{0.604} & 0.522 & 0.478 \\

\textbf{Ours} & \textbf{0.782} & \textbf{0.640} & \textbf{0.632} & \textbf{0.510} & 0.588 & \textbf{0.525} & 0.510 \\
\hline
 & telephone & cabinet & loudspeaker & watercraft & lamp & rifle & mean \\
\hline
Retrieval & 0.622 & 0.518 & 0.468 & 0.422 & 0.325 & 0.475 & 0.455 \\

Auto-encoder & 0.706 & 0.663 & 0.629 & 0.556 & 0.431 & 0.605 & 0.582 \\

Sketch2Model & 0.719 & \textbf{0.701} & \textbf{0.641} & \textbf{0.586} & \textbf{0.472} & 0.612 & 0.601 \\

\textbf{Ours} & \textbf{0.757} & 0.699 & 0.630 & 0.583 & 0.466 & \textbf{0.632} & \textbf{0.611} \\ 
\hline\hline
\end{tabular}}
\end{center}
\label{table:table1}

\end{table*}

\vspace*{1\baselineskip}
\noindent\textit{3.3. Loss Function}
\vspace*{1\baselineskip}

To effectively train the network, the loss functions are thoughtfully designed and consist of three key components: 1) a multi-scale mIoU loss $\mathcal{L}_{sp}$, 2) flatten loss and laplacian smooth loss $\mathcal{L}_{r}$, and 3) a structure-aware GAN loss $\mathcal{L}_{sd}$.
The multi-scale mIoU loss $\mathcal{L}_{sp}$evaluates the correspondence between rendered silhouettes and ground truth silhouettes by quantifying their similarity. To improve the computational efficiency, we incrementally enhance the resolutions of silhouettes, which is represented as: 
\begin{align}
\mathcal{L}_{s p}=\sum_{i=1}^{N} \lambda_{s_{i}} \mathcal{L}_{iou}^{i}
\label{lsp}
\end{align}

\noindent$\mathcal{L}_{iou}$ is defined as:
\begin{align}
\mathcal{L}_{i o u}\left(S_{1}, S_{2}\right)=1-\frac{\left\|S_{1} \otimes S_{2}\right\|_{1}}{\left\|S_{1} \oplus S_{2}-S_{1} \otimes S_{2}\right\|_{1}}
\end{align}
where $S_1$ and $S_2$ is the rendered silhouette. 

We introduced the adoption of flatten loss and Laplacian smooth loss to enhance the realism and visual quality of the generated meshes, represented by $\mathcal{L}_{r}$, as shown in~\cite{zhang2021sketch2model,kato2018neural,liu2019soft}.

For our structure-aware GAN loss  $\mathcal{L}_{sd}$, non-saturating GAN loss \cite{mescheder2018training} is used.
\begin{align}
\begin{split}
\mathcal{L}_{sd} &=\mathbf{E}_{\mathbf{z_v} \sim p_{z_v}, \xi \sim p_{\xi}}\left[f\left(SD\left(R(M, \xi)\right)\right)\right] \\
&+\mathbf{E}_{\mathbf{z_{vr}} \sim p_{z_{vr}}, \xi \sim p_{\xi}}\left[f\left(-SD(R(M_r, \xi))\right)\right] \label{gan}
\end{split}\\ 
& \mathit{{ where  }}  f(u)=-\log (1+\exp (-u))
\end{align}

The overall loss function $Loss$ is calculated as the weighted sum of the three components:
\begin{align}
Loss =  \mathcal{L}_{sp} + \mathcal{L}_{r} + \lambda_{sd} \mathcal{L}_{sd}
\label{loss}
\end{align}

\section{Experiments and results}

\vspace*{1\baselineskip}
\noindent\textit{4.1. Datasets}
\vspace*{1\baselineskip}

It is uncommon to find publicly available datasets that contain both sketches and their corresponding 3D models, making them quite scarce in availability. In line with the approach described in \cite{zhang2021sketch2model},we opted for an alternative strategy by utilizing synthetic data from ShapeNet-synthetic for training purposes. Subsequently, we evaluated the performance of the trained network on real-world data from ShapeNet-sketch. The synthetic data was obtained by extracting edge maps using a canny edge detector on rendered images from Kar et al. [48], featuring 13 categories of 3D objects. The ShapeNet-Sketch dataset was drawn by human volunteers, with varying skill levels, who draw objects based on images of 3D objects from [48]. The dataset includes a total of 1300 sketches and their corresponding 3D shapes.

\begin{table*}[width=.72\textwidth, cols=4, pos=h]
\caption{The quantitative evaluation of ShapeNet-Sketch dataset.}
\begin{center}{
\begin{tabular}{c|ccccccc}
\hline\hline
\multicolumn{8}{c}{Shapenet-sketch (Voxel IoU $\uparrow$)} \\
\hline
 & car & sofa & airplane & bench & display & chair & table \\
\hline
Retrieval & 0.626 & 0.431 & 0.411 & 0.219 & 0.338 & 0.238 & 0.232 \\

Auto-encoder & 0.648 & \textbf{0.534} & 0.469 & 0.347 & 0.472 & 0.361 & 0.359 \\

Sketch2Model  & 0.659 & \textbf{0.534} & 0.487 & 0.366 &  \textbf{0.479} &  \textbf{0.393} & 0.357\\

\textbf{Ours} &  \textbf{0.675} &  \textbf{0.534} &  \textbf{0.490} & \textbf{0.368} & 0.463 & 0.382 &  \textbf{0.370} \\
\hline
 & telephone & cabinet & loudspeaker & watercraft & lamp & rifle & mean \\
\hline
Retrieval & 0.536 & 0.431 & 0.365 & 0.369 & 0.223 & 0.413 & 0.370 \\

Auto-encoder & 0.537 & 0.534 &  0.533 & 0.456 & 0.328 & 0.541 & 0.372 \\

Sketch2Model & 0.554 & \textbf{0.568} & \textbf{0.544} & 0.450 & 0.338 & 0.534 & \textbf{0.483} \\

\textbf{Ours} & \textbf{0.576} & 0.553 & 0.514 & \textbf{0.467} &  \textbf{0.347} &  \textbf{0.543} &  \textbf{0.483} \\
\hline\hline
\end{tabular}}
\end{center}
\label{table:table2}

\end{table*}

\vspace*{1\baselineskip}
\noindent\textit{4.2. Implementation Details}
\vspace*{1\baselineskip}

for the encoder $E$, we utilized a ResNet-18 architecture \cite{he2016deep}   to extract image features in our approach. To render silhouettes, we employed the SoftRas algorithm \cite{liu2019soft}. For the canonical view, the 3D objects were uniformly oriented with an elevation angle of 0 degrees and an azimuth angle of 0 degrees. Moreover, they were consistently positioned at a constant distance from the camera. The ground-truth viewpoint was utilized for the rendering process. For each predicted model and GT model, we used N=2 for rendering. The silhouette with the corresponding ground truth was used for calculating the IoU loss. The Adam optimizer was employed with an initial learning rate of 1e-4 and multiplied by 0.3 every 800 epochs. The beta values were set to 0.9 and 0.999. The total training epochs were set to 2000. The model was trained individually for each class of the dataset. The value of $\lambda$ in Equation. \ref{pam} was set to 1 and the value of $\lambda_{sd}$ in Equation. \ref{loss} was set to 0.1. The camera poses used for rendering the silhouettes were sampled from a uniform distribution. The model was trained and evaluated on four NVIDIA GeForce RTX3090 GPUs.
\begin{table*}[width=.85\textwidth, cols=4, pos=h]
\begin{center}
\caption{Robustness test for partial sketches.}
{
\begin{tabular}{c|c|ccccccc}
\hline\hline
\multicolumn{9}{c}{Shapenet-synthetic (Voxel IoU $\uparrow)$} \\

\hline
                &       & airplane       & bench          & cabinet        & car            & chair          & lamp           & rifle          \\ \hline
Original Input & Sketch2model & 0.624          & 0.481          & \textbf{0.701} & 0.751          & 0.522          & \textbf{0.472} & 0.612          \\
                    & \textbf{Ours}         & \textbf{0.632} & \textbf{0.510} & 0.692          & \textbf{0.782} & \textbf{0.526} & 0.467          & \textbf{0.631} \\ 
10$\%$ Missing & Sketch2model & 0.560          & 0.454          & \textbf{0.675} & 0.695          & 0.483          & \textbf{0.445} & 0.565          \\
                    & \textbf{Ours}         & \textbf{0.578} & \textbf{0.476} & 0.663          & \textbf{0.743} & \textbf{0.486} & 0.442          & \textbf{0.595} \\ 
20$\%$ Missing & Sketch2model & 0.481          & 0.419          & \textbf{0.638} & 0.619          & \textbf{0.434} & \textbf{0.403} & 0.505          \\
                    & \textbf{Ours}         & \textbf{0.506} & \textbf{0.438} & 0.635          & \textbf{0.689} & 0.432          & 0.402          & \textbf{0.543} \\ \hline
                &          & sofa           & table          & telephone      & watercraft     & diplay         & loudspeaker    & mean           \\ \hline
Original Input & Sketch2model & 0.622          & 0.478          & 0.719          & \textbf{0.586} & \textbf{0.604} & \textbf{0.641} & 0.601          \\
                    & \textbf{Ours}         & \textbf{0.640} & \textbf{0.510} & \textbf{0.758} & 0.584          & 0.589          & 0.631          & \textbf{0.611} \\ 
10$\%$ Missing & Sketch2model & 0.580          & 0.437          & 0.704          & \textbf{0.549} & \textbf{0.573} & \textbf{0.628} & 0.565          \\
                    & \textbf{Ours}         & \textbf{0.604} & \textbf{0.465} & \textbf{0.733} & 0.547          & 0.552          & 0.614          & \textbf{0.576} \\ 
20$\%$ Missing & Sketch2model & 0.516          & 0.400          & 0.682          & 0.489          & \textbf{0.525} & \textbf{0.606} & 0.517          \\
                    & \textbf{Ours}         & \textbf{0.550} & \textbf{0.425} & \textbf{0.711} & \textbf{0.499} & 0.499          & 0.595          & \textbf{0.532} \\ \hline\hline
\end{tabular}}
\end{center}

\label{table:tablevii}

\end{table*}

\vspace*{1\baselineskip}
\noindent\textit{4.3. Results}
\vspace*{1\baselineskip}

\textit{\textbf{1) The ShapeNet-Synthetic Dataset:}} We conducted a comparative analysis between our method and the model retrieval approach, utilizing features extracted from a pretrained network specialized in sketch classification. Additionally, we benchmarked our approach against the current state-of-the-art (SOTA) model, following the experimental setup outlined in \cite{zhang2021sketch2model}. We first tested it on the ShapeNet-Synthetic dataset, which provided accurate ground truth 3D models for training and evaluation. The voxel IoU metric, a commonly used measure for 3D reconstruction, was employed to assess the fidelity of the generated meshes, as shown in Table \ref{table:table1}. Our approach was subjected to a quantitative evaluation, which highlighted its effectiveness by achieving state-of-the-art (SOTA) performance. Additionally, we compared our method with existing state-of-the-art techniques through a quantitative evaluation. The results further demonstrated the superiority of our approach in generating models with enhanced quality and fidelity in terms of structure. These findings are visually represented in Fig. \ref{fig:2}.

\textit{\textbf{2) The ShapeNet-Sketch Dataset:}} 2)Our approach is further validated through evaluating the performance of our
model on real-world human drawings, which are more challenging due to the varied drawing skills and styles of the creators. Domain gaps exist between the synthetic data used for training (ShapeNet-Synthetic) and the real data used for evaluation (ShapeNet-Sketch). To address this challenge, we developed a powerful and robust feature extractor with structural awareness. Our experiments demonstrated that our model generalized well on real data, indicating its ability to overcome the domain gaps and perform effectively in real-world scenarios. The results, shown in Table \ref{table:table2}, demonstrate that our model outperforms existing state-of-the-art models in most categories, highlighting the effectiveness of our approach.It is worth noting that the potential of domain adaptation techniques could be explored as a means to enhance the network's performance in real datasets with domain gaps. This approach has the potential to bridge the gap between synthetic and real data and further improve the network's performance in real-world scenarios. This avenue could be considered for future research and development in the field.

\vspace*{1\baselineskip}
\noindent\textit{4.4. Evaluating Runtime for 3D Modeling}
\vspace*{1\baselineskip}



\begin{table}[width=.50\textwidth, cols=4, pos=h]
\setlength{\tabcolsep}{4mm}
\caption{Average runtime for generatin a single 3D model.}
\begin{center}
\scalebox{0.82}{
\begin{tabular*}{1.2\linewidth}{ c | c | c | c}
\hline\hline
 \textbf{Inference by GPU} & 0.011 s & \textbf{Inference by CPU} & 0.062 s \\
\hline\hline
\end{tabular*}}
\end{center}
\label{table:table3}

\end{table}
After training our network, We assessed its performance on a personal computer that was equipped with a consumer-grade graphics card, specifically the NVIDIA GeForce RTX 3090 model. The results, as shown in Table 4,demonstrated that our method achieved a generation speed of 90FPS. This represents a 38.9$\%$ speed improvement compared to Sketch2Model (55FPS) \cite{zhang2021sketch2model}. Furthermore, we tested the performance of our method solely on the CPU (Intel Xeon Gold 5218) and observed an 11.4$\%$ speed gain compared to Sketch2Model (14FPS) \cite{zhang2021sketch2model}. Our network achieved a rate of 16FPS, which is sufficient for smooth computer-human interaction. These findings highlight the efficiency of our approach in enabling rapid 3D modeling, regardless of the hardware configuration used.

\vspace*{1\baselineskip}
\noindent\textit{4.5. Robustness to Partial Input}
\vspace*{1\baselineskip}

In some industrial applications, for example, in dirty environments and in cases of continual use, drawing continuous line using touch screens may be challenging \cite{greenstein1997pointing}. In this section, we demonstrate that our approach is robust to partial output of the sketch -- the user draw sketch is missing in some parts (with broken/unconnected lines). We created a random-sized and randomly positioned blank rectangle mask to obscure parts of the original sketch, creating a corrupted sketches. This corrupted sketch was added to the database if the difference between the original sketch and the corrupted sketch falls within a range of 10-20$\%$, simulating the typical missing content found in sketches. We use corrupted sketches at the inference stage, and the result is shown in Table 3. 

The result shows that our method, with the involvement in Shape Discriminator (SD) and Stroke Enhancement Module (SEM), can successfully handle the corrupted sketches without significant loss in performance, demonstrating high robustness in real-world industrial applications. 

\vspace*{1\baselineskip}
\noindent\textit{4.6. Manufacturing the Generated Models}
\vspace*{1\baselineskip}

\begin{figure}[h]
\centering
\includegraphics[width=0.95\linewidth]{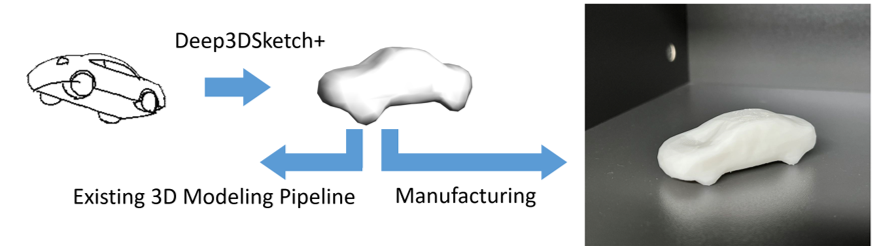}
\caption{{\textbf{An example of the application of Deep3DSketch+ to produce customized models.} Users can use a single sketch to generate a model that can be either feed into existing 3D modeling pipeline or directly manufacture it to be a real object.}}
\label{fig:9}
\end{figure}

Our approach is designed to produce watertight 3D models that can easily be integrated into existing manufacturing pipelines. To demonstrate this, we show an example of a physical model produced by a consumer-level 3D printer (KOKONI EC-1, Moxin Technology) using a sketch-based 3D modeling result and the default cloud-based slicer to generate the toolpath. The generated model can also be exported directly to existing 3D modeling tools for editing or be used as the template for more creations (Fig. \ref{fig:9}). The example illustrates that our approach can be easily adapted to current manufacturing processes.

\noindent\textit{4.7. Ablation Study}
\vspace*{1\baselineskip}

\begin{figure}[h]
\centering
\resizebox{.43\textwidth}{!}{
\includegraphics[width=0.43\linewidth]{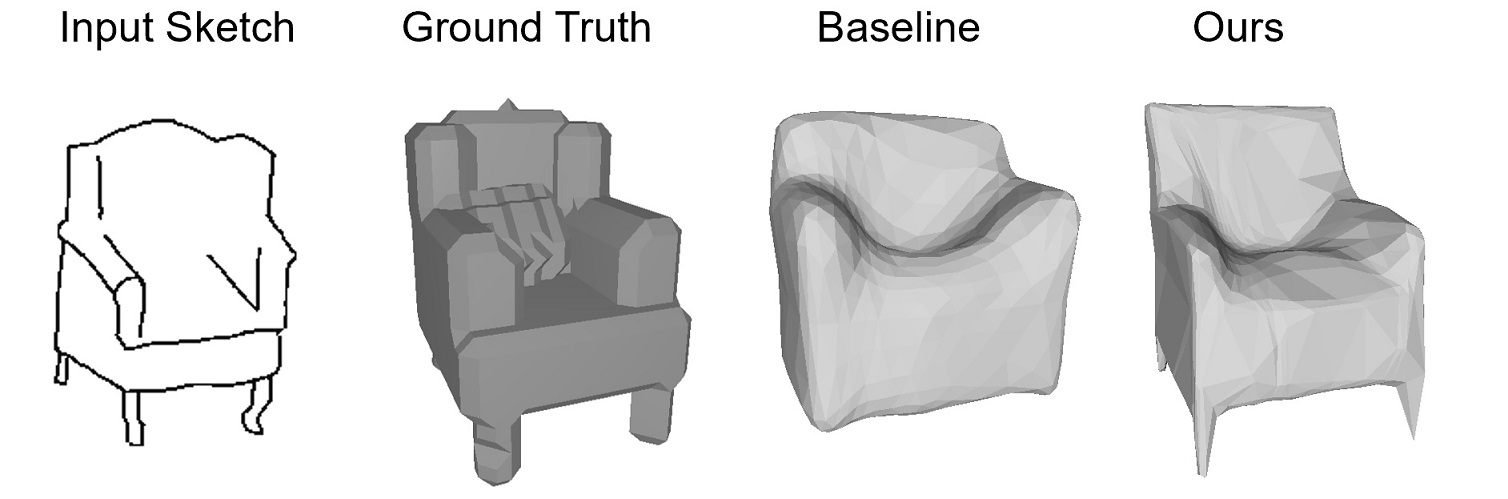}}
\caption{{\textbf{Ablation study.} Our method generates more fine-grained structures compared to the baseline method.}}
\label{fig:4}
\vspace{0.1cm}
\end{figure}

To validate the efficacy of our proposed methodology, we conducted an ablation study as presented in Table \ref{table:table3}. The results demonstrated that the inclusion of the Shape Discriminator (SD) and Stroke Enhancement Module (SEM) in our method led to performance improvements. These enhancements contributed to the production of 3D models with higher fidelity, as illustrated in Fig. \ref{fig:4}.  In comparison to the baseline method without SD or SEM, our method showcased superior performance and generated models with enhanced quality and fidelity.

\begin{table}[width=.72\textwidth, cols=4, pos=h]
\caption{Ablation study.}
\begin{center}
\scalebox{0.67}{
\begin{tabular*}{1.5\linewidth}{c c | c c c c c c c }
\hline\hline
SD & SEM & car & sofa & airplane & bench & display & chair & table \\
\hline
 & & 0.767  & 0.630  & 0.633 & 0.503 & 0.586 & 0.524  & 0.493  \\
$\surd$ & & 0.778  & 0.632 & \textbf{0.637} & 0.503 & \textbf{0.588} & 0.523  & 0.485  \\
$\surd$ & $\surd$ & \textbf{0.782}  & \textbf{0.640}  & 0.632 & \textbf{0.510}  & \textbf{0.588}  & \textbf{0.525}  & \textbf{0.510}  \\
\hline
SD & SEM & telephone & cabinet & loudspeaker & watercraft & lamp & rifle & mean \\
\hline
 & & 0.742 & 0.690 & 0.555 & 0.563 & 0.458 & 0.613 & 0.598 \\
$\surd$ & & 0.749  & 0.688  & 0.617 & 0.567 & 0.454 & 0.612  & 0.602  \\
$\surd$ & $\surd$ & \textbf{0.757}  & \textbf{0.699} & \textbf{0.630}  & \textbf{0.583}  & \textbf{0.466} & \textbf{0.624}  & \textbf{0.611}  \\
\hline\hline
\end{tabular*}}
\end{center}
\label{table:table3}

\end{table}

\begin{figure}[h]
\centering
\includegraphics[width=0.95\linewidth]{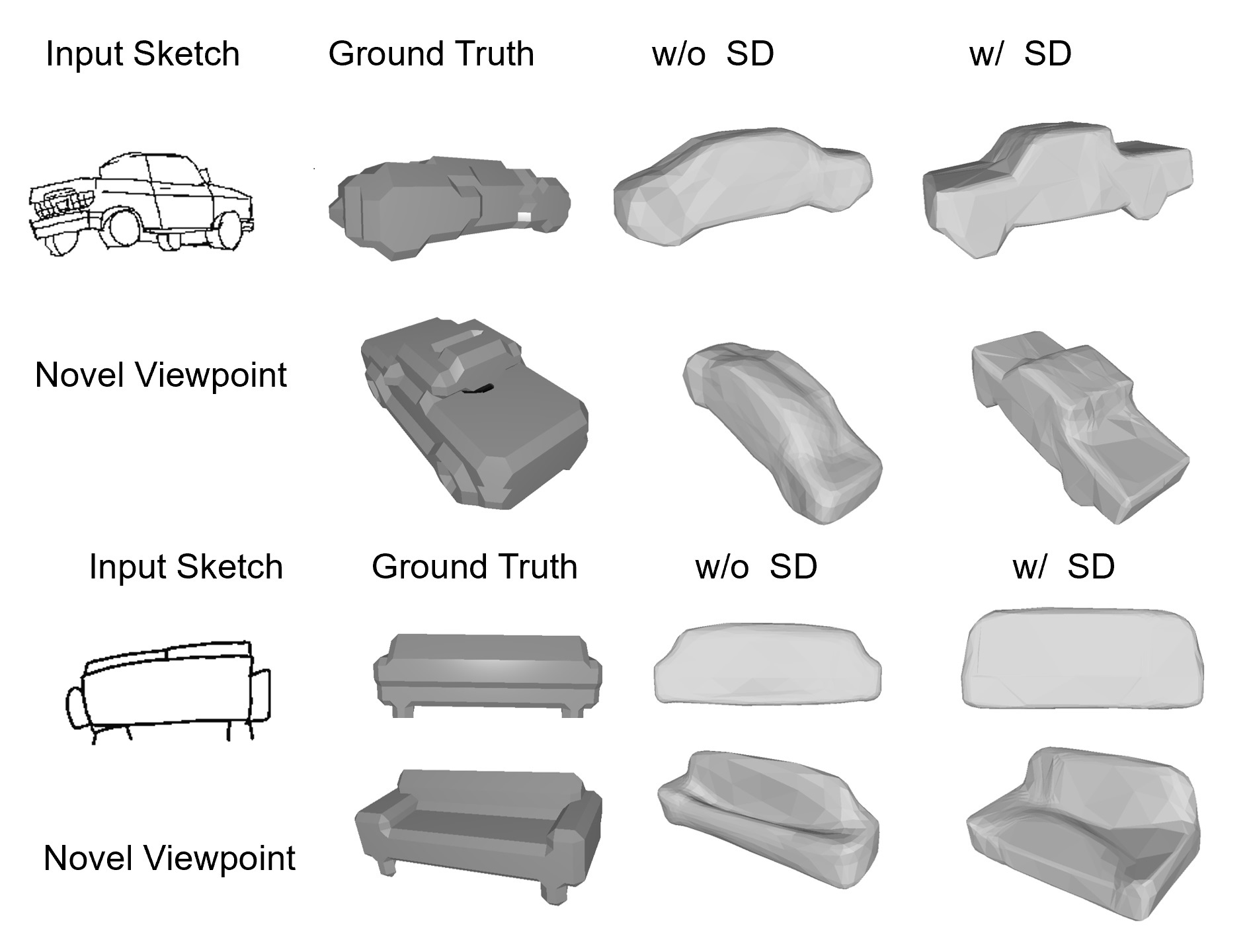}
\caption{{\textbf{The effectiveness of SD and random-viewpoint sampling.} As shown in the example, the neural network generates more fine-grained structures compared to the baseline method.}}
\label{fig:5}
\end{figure}

Moreover, we argue that our proposed method of using a Shape Discriminator (SD) and random viewpoint sampling allows the neural network to learn from real shapes from multiple angles, resulting in the ability to predict reasonable structural information that is not present in the sketch. This is evident in the examples shown in Fig. \ref{fig:5}, where the sitting pad on a sofa is reconstructed even though the input sketch is only viewed from the back, and the flat plane at the back of a car is reconstructed even though the input sketch is only viewed from the front. This demonstrates the effectiveness of our approach in generating high-quality 3D models.

\section{Conclusion}
\vspace*{1\baselineskip}
Our approach, Deep3DSketch+, introduce a new way of creating 3D models using free-hand sketches as input. Traditional CAD-based modeling methods can be time-consuming and complex, but our method offers a more intuitive and efficient solution. By utilizing a neural network with a Shape Discriminator (SD) and Stroke Enhancement Module (SEM), we are able to overcome the challenges of sparse and ambiguous sketches. The designed algorithm achieves the state-of-the-art (SOTA) performance at both synthetic and real-world data. The algorithm is also robust to partial sketch input in industrial applications. Additionally, our generated models are watertight and ready for integration into existing manufacturing processes. We have also demonstrated the ability to turn a free-hand sketch into a physical 3D object through additive manufacturing. We believe that our approach has great potential for more rapid product designs and democratizing manufacturing.

\section*{Declaration of competing interest}
The authors declare that they have no known competing financial interests or personal relationships that could have appeared
to influence the work reported in this paper.

\section*{Data availability}
The data that support the findings of this study are available on request from the corresponding author [T. Chen], upon reasonable request.

\section*{Acknowledgments}
This work was supported in part by National Natural Science Foundation of China (No. 61772198, No. U20A20228), and Zhejiang Province Key Laboratory of Smart Management and Application of Modern Agricultural Resources (No. 2020E10017). 

\bibliographystyle{model1-num-names}
\bibliography{cas-refs}

\end{document}